\documentclass[lettersize,journal]{IEEEtran}
\usepackage{amsmath,amsfonts}
\usepackage[linesnumbered,ruled,vlined]{algorithm2e}
\usepackage{amsthm}
\usepackage{array}
\usepackage{textcomp}
\usepackage{stfloats}
\usepackage{url}
\usepackage{verbatim}
\usepackage{graphicx}
\usepackage{cite}
\begin{document}

\title{Decision Transformers for RIS-Assisted Systems with Diffusion Model-Based Channel Acquisition}

\author{Jie Zhang, Yiyang Ni, Jun Li, Guangji Chen, Zhe Wang, Long Shi, Shi Jin, Wen Chen, and H. Vincent Poor
        \IEEEcompsocitemizethanks{
        \IEEEcompsocthanksitem Part of this work has been presented in IEEE ICCC 2024 [DOI: 10.1109/ICCC62479.2024.10681714].
        
        Jie Zhang, Guangji Chen and Long Shi are with the School of Electronic and Optical Engineering, Nanjing University of Science and Technology, Nanjing 210094, China (e-mail: zhangjie666@njust.edu.cn, guangjichen@njust.edu.cn, slong1007@gmail.com).
        \IEEEcompsocthanksitem Yiyang Ni is with the Jiangsu Second Normal University and Jiangsu Institute of Educational Science Research, Nanjing 210094, China (email: niyy@jssnu.edu.cn). \textit{(corresponding author)}.
        \IEEEcompsocthanksitem Jun Li and Shi Jin are with the National Mobile Communications Research Laboratory, Southeast University, China (jleesr80@gmail.com, jinshi@seu.edu.cn).
        \IEEEcompsocthanksitem Zhe Wang is with the School of Computer Science and Engineering, Nanjing University of Science and Technology, Nanjing 210094, China (email: zwang@njust.edu.cn).
        \IEEEcompsocthanksitem Wen Chen is with the Department of Electronic Engineering, Shanghai Jiao Tong University, Shanghai 200240, China (e-mail: wenchen@sjtu.edu.cn).
        \IEEEcompsocthanksitem H. Vincent Poor is with the Department of Electrical and Computer Engineering, Princeton University, Princeton, NJ 08544, USA (poor@princeton.edu).
        }
        }
\maketitle

\begin{abstract}
Reconfigurable intelligent surfaces (RISs) have been recognized as a revolutionary technology for future wireless networks. However, RIS-assisted communications have to continuously tune phase-shifts relying on accurate channel state information (CSI) that is generally difficult to obtain due to the large number of RIS channels. The joint design of CSI acquisition and subsection RIS phase-shifts remains a significant challenge in dynamic environments. In this paper, we propose a diffusion-enhanced decision Transformer (DEDT) framework consisting of a diffusion model (DM) designed for efficient CSI acquisition and a decision Transformer (DT) utilized for phase-shift optimizations. Specifically, we first propose a novel DM mechanism, i.e., conditional imputation based on denoising diffusion probabilistic model, for rapidly acquiring real-time full CSI by exploiting the spatial correlations inherent in wireless channels. Then, we optimize beamforming schemes based on the DT architecture, which pre-trains on historical environments to establish a robust policy model. Next, we incorporate a fine-tuning mechanism to ensure rapid beamforming adaptation to new environments, eliminating the retraining process that is imperative in conventional reinforcement learning (RL) methods. Simulation results demonstrate that DEDT can enhance efficiency and adaptability of RIS-aided communications with fluctuating channel conditions compared to state-of-the-art RL methods.
\end{abstract}


\section{Introduction}
Reconfigurable intelligent surfaces (RISs) have emerged as a revolutionary technology in wireless communications for enhancing coverage and capacity. Numerous passive reflecting elements in an RIS are capable of adjusting phase shift of incident signals dynamically over time, thereby intelligently reshaping wireless environments~\cite{IRS_technology_4,IRS_technology_2,IRS_technology_5,IRS_ZJ_MARL, CGJ_IRS_1}. By precisely controlling the phase of these reflecting elements, an RIS can coherently superimpose or cancel reflecting signals, enhancing or attenuating the signal strength at the receiver~\cite{IRS_construction_1,IRS_construction_2,CGJ_IRS,IRS_construction_3,IRS_ZJ_how_often, CGJ_IRS_2}. This characteristic endows RISs with the ability to achieve higher transmission rates, greater energy saving, and less latency. Despite these advantages, the effectiveness of RISs is heavily reliant on the ability to accurately manage channel conditions. To fully realize these benefits, the key challenges remain in accurately acquiring the channel state information (CSI) and efficiently optimizing beamforming strategies for adapting to dynamic channel conditions.

The passive nature of RISs makes channel estimation more complicated than in conventional communication systems. Some studies~\cite{IRS_perfect_CSI}, for analytical convenience, bypass channel estimation and assume that the CSI is fully known and can be directly acquired. However, this assumption is generally strong in practical systems. 
Unlike traditional multi-antenna communication systems, a fully passive RIS lacks the ability to sense or receive pilot signals directly, which makes direct channel estimation unfeasible. Instead, it is more appealing to estimate the cascaded channels that pass from the BS to the RIS and subsequently to users. For instance, the cascaded channels can be estimated successively by switching the on/off states of RIS elements~\cite{IRS_mod_on_off}. Channel estimation techniques, such as least square~\cite{IRS_CE_LS} and minimum mean-square error~\cite{IRS_CE_MSE}, can be applied to cascaded channel estimation, while they often require prior statistical information, such as mean values and covariance matrices that are difficult to obtain. 

To address this issue, recent advancements in artificial intelligence (AI)~\cite{BLADE_FL_SYM, UAV_LiuQian, FL_ZSY} have led to data-driven deep learning approaches that can effectively capture complex patterns, providing an alternative to traditional model-based methods that rely on prior knowledge. The work in~\cite{IRS_CNN} employed a convolutional neural network (CNN) for mapping pilot signals to channels, enabling CSI prediction for each element via a CNN estimator. The work in~\cite{IRS_GAN} proposed a conditional generative adversarial network (GAN) method to estimate cascaded channels using received signals as conditional information, where the adversarial training between the generator and discriminator networks ensured the generated channels closely resembled the actual channels. Similarly, a generative diffusion model was employed in~\cite{IRS_DM}, where the channel acquisition process was considered as a denoising task, using pilot symbols as auxiliary inputs to progressively recover the true channel from an artificial Gaussian noise. 

Despite their potential, these methods face significant challenges, particularly with high training costs. In a large-scale RIS system with massive reflecting elements, the pilot overhead for channel estimation scales with the number of elements, which substantially reduces the available time resources for subsequent data transmission. The element-grouping methods in~\cite{IRS_Group_Based} significantly reduced the number of required pilot symbols by grouping RIS elements and leveraging RIS symmetry. Although this method helps alleviate the burden of pilot overhead, grouped estimation leads to inaccurate channel estimation, thereby degrading the communication performance due to the induced lower passive beamforming gain.

Apart from the CSI acquisition issue, another significant challenge is the stochastic optimization of RIS beamforming. Conventional optimization methods, such as the semi-definite relaxation (SDR) method and alternating optimization (AO) approach~\cite{IRS_AO_2} typically rely on complex iterative computations like cyclic high-dimensional matrix calculations. This dependency not only increases the computational burden but also limits their applicability in dynamic communication environments. To address this issue, model-free reinforcement learning (RL) approaches have been deployed to optimize the beamforming decision-making process. In~\cite{IRS_DQN_1,IRS_DQN_2,IRS_DQN_3}, the deep Q-network (DQN) algorithm was used to optimize RIS discrete phase shifts to maximize system throughput and eliminating the need for intricate mathematical models. Similarly, the deep deterministic policy gradient (DDPG) algorithm was applied in~\cite{IRS_DDPG_1} and ~\cite{IRS_DDPG_2}, considering estimated channel states, previous phase shifts, and received SINR to optimize RIS phases and maximize the sum rate in a non-orthogonal multiple access (NOMA) system. The proximal policy optimization (PPO) algorithm in~\cite{IRS_PPO} jointly optimized transmit beamforming and phase shift design, aiming to improve system capacity in an RIS-aided integrated sensing and communication system.

Despite the significant potential of RL methods in RIS-assisted communication environments, they struggle to generalize across varied scenarios. Changes in dynamic environments affecting CSI distributions will inevitably hinder the performance of RL models, requiring resource-intensive retraining of RL from the scratch when adapting to new scenarios. Under these circumstances, conventional RL algorithms struggle to effectively apply beamforming policies directly to new scenarios, which necessitates extra retraining and thus consuming substantial computational resources. 

To improve adaptability and generalization, the decision Transformer (DT) architecture~\cite{decision_transformer} represents a notable advancement. The DT architecture leverages a sequence-based modeling approach, drawing on the robust representation and generalization capabilities of generative pre-trained Transformer (GPT) architectures to tackle complex decision-making tasks. Unlike conventional RL algorithms, DTs can learn from historical decision sequences, directly mapping to optimal beamforming strategies without requiring retraining from the ground up, which significantly enhance the adaptability in new scenarios. By minimizing reliance on repetitive, resource-intensive calculations and leveraging pre-trained models for rapid fine-tuning, DTs provide a scalable and efficient framework for managing RIS-assisted communication systems~\cite{IRS_ZJ_DT}.



To tackle the aforementioned challenges, this paper proposes a diffusion-enhanced decision Transformer (DEDT) framework orchestrating the strengths of both DM and DT, to enable efficient channel estimation and adaptive beamforming in RIS-assisted communication systems. To the best of the authors' knowledge, our work is the first of its kind that introduce DT frameworks, combined with DM, to wireless communications for decision making. Towards a robust solution for dynamic RIS environments, the proposed framework not only boosts channel estimation accuracy and reduces the computational overhead for CSI acquisition, but also saves a huge amount of resources for RL model's online retraining. Our main contributions can be summarized as follows.

\begin{itemize}
    \item Our core contribution is the development of a sophisticated channel estimation method that leverages the spatial correlation among RIS elements. Unlike conventional methods that require direct estimation of each element, our approach estimates the CSI for a selected subset of reflecting elements and extrapolates this information to infer the CSI across the entire surface. This significantly reduces the computational demands and operational costs. By enhancing the channel estimation accuracy and efficiency, our strategy supports precise beamforming, ultimately improving both the quality and capacity of wireless communication systems.

    \item We propose a DT framework to address the challenges of dynamic and unpredictable communication environments. Our framework employs Transformer architectures renowned for their robust representation and generalization capabilities to map historical data sequences to optimal beamforming configurations, and thus facilitate dynamic adaptations of beamforming strategies without the need for extensive retraining. The ability of our DT framework to adapt to new scenarios greatly enhances the flexibility and responsiveness of the RIS-assisted systems, relative to conventional RL methods.

    \item Our DEDT architecture incorporates DM with DT frameworks to refine the process of generating CSI predictions, and provide real-time, accurate predictions. This integration enables precise and timely adjustments to RIS configurations, thereby not only boosting the accuracy of the system responses to varying communication environments, but also reducing the latency typically associated with such adaptations.
\end{itemize}

Our paper is organized as follows. Section II provides background information on DM and DT. In Section III, we introduce the system model and formulate the optimization problem.  Section IV describes the process of obtaining the channel with DM and generalizing the policy using DT. Section V presents the DEDT method for the offline training and online inference. Section VI analyzes the simulation results, and Section VII concludes the paper.

\emph{Notation:} Vectors and matrices are denoted by boldface lowercase and uppercase letters, respectively. The operations $(\cdot)^{T}$ and $(\cdot)^{H}$ represent the transpose and conjugate transpose operations, respectively. For a vector $\boldsymbol{x}$, $\boldsymbol{x}^{n}$, $\left\|\boldsymbol{x}\right\|$ and $\mathrm{diag}(\boldsymbol{x})$ denote the $n$-th element, the Euclidean norm, and the diagonal matrix with the elements of $\boldsymbol{x}$, respectively. For a matrix $\boldsymbol{A}$, the element in the $i$-th row and $j$-th column is denoted by $\boldsymbol{A}^{ij}$. $\mathbb{C}$ and $\mathbb{R}$ stand for the complex number domain and the real number domain, respectively. The complex Gaussian distribution with mean $\mu$ and variance $\sigma^{2}$ is denoted by $\mathcal{CN}(\mu,\sigma^{2})$. $\mathbb{E}[\cdot]$ denotes the statistical expectation. The main symbol notations used in this paper are summarized in Table~\ref{tab:sum_nota}.

\begin{table}[ht]\caption{Summary of main notation}
\centering
\begin{tabular}{l l}
\hline
\textbf{Notation} & \textbf{Description}\\
\hline
$M$, $N$ & Number of BS antennas, number of RIS elements\\
$\boldsymbol{\mathrm{G}}_{t}$ & Channel state information between BS and RIS\\
$\boldsymbol{\mathrm{h}}_{t}$ & Channel state information between RIS and user\\
$\boldsymbol{\mathrm{\Phi}}_{t}$, $\phi_{t}$ & Beamforming matrix, beamforming vector\\
$\boldsymbol{x}_{t}^{(k)}$ & Latent data at diffusion step $k$\\
$b_{k}$ & The variance of the diffusion noise\\
$\tau_{t}$ & The sample sequence\\
$\hat{R}_{t}$ & Future cumulative rewards\\
\hline
\end{tabular}
\label{tab:sum_nota}
\end{table}

\section{Preliminary}
In this section, we will introduce some background knowledge about DM and DT.
\subsection{Diffusion Model}
\begin{figure}[h]
    \centering
    \includegraphics[width=0.45\textwidth]{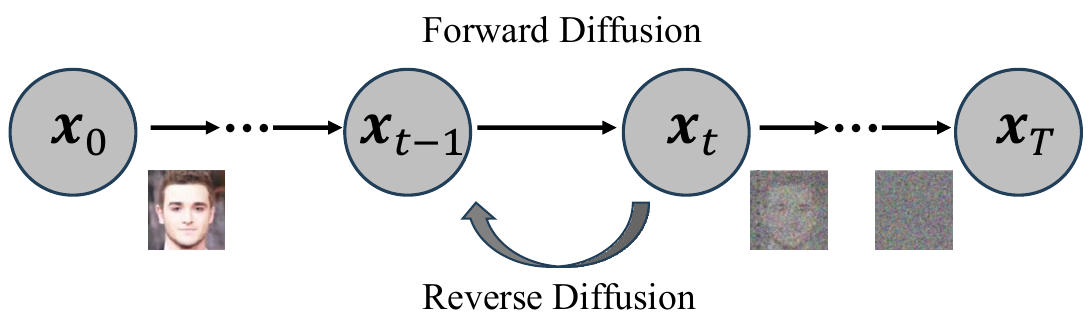}
    \caption{Illustration of diffusion model structure.}
	\label{diffusion model}
\end{figure}

DMs represent a class of generative neural networks that draw inspiration from the principles of statistical physics, particularly the concept of diffusion processes. At their core, these models aim to simulate a process that incrementally transforms complex data distributions into simple, well-understood noise distributions (typically Gaussian distribution) and then learns how to reverse this transformation to generate new samples. This dual-phase process enables the generation of high-quality, diverse data, making diffusion models an exciting alternative to traditional generative models like GAN architecture. As shown in 
Fig.~\ref{diffusion model}~\cite{DDPM}, the operation of diffusion models can be broken down into two key phases: The forward diffusion phase and the reverse diffusion phase. 

In the forward diffusion phase, the model gradually corrupts the original data by adding Gaussian noise over a sequence of discrete diffusion steps. This process can be likened to observing an object through a progressively foggier over time, and the object becomes increasingly obscured until it eventually blends into noises. The forward diffusion follows a Markov chain, where each step of adding noise depends only on the previous state rather than on earlier diffusion steps. The corruption process continues until the original data distribution is almost indistinguishable from pure Gaussian noise. At each diffusion step, artificial noise is added in such a way that the data distribution becomes progressively entropic, eventually resulting in a high-entropy, noise-like state. This phase is designed such that the transition probabilities are known and do not require learning, simplifying the modeling process.

The reverse phase is where the generative capability of the diffusion model comes into play. Here, the model aims to learn the inverse of the forward process starting from pure noise, it iteratively denoises the data, removing the noise added at each previous step to reconstruct the original data distribution. The goal of the reverse phase is to generate a sequence of denoised data states that eventually resemble the original training data. The reverse diffusion process is trained to predict the noise introduced at each step of the forward process. This is achieved through a neural network that learns to estimate the noise at each diffusion step and subtract it accordingly. Unlike the forward phase, the reverse process is not predetermined but is learned through training on data samples, making it a critical part of the model's generative capability.

One of the key strengths of DMs lies in their ability to generate high-quality samples. This capability comes from their methodical approach to generation, which allows for finer control over the denoising process, ultimately leading to more stable and diverse outputs compared to some other generative methods. Additionally, DMs exhibit robustness when applied to sequential data generation, such as audio synthesis and time-series data. The structured approach of gradually adding and then removing noise allows the model to maintain temporal and spatial dependencies across sequences, capturing subtle patterns that are essential in generation tasks. This makes DMs especially valuable in fields where traditional generative methods might struggle to maintain the coherence of intricate temporal and spatial structures.

\subsection{Decision Transformer}
DTs represent a novel approach to tackling RL tasks by framing them as sequence modeling problems, inspired by advancements in natural language processing (NLP) and computer vision using Transformer architectures. In conventional RL algorithms, the goal is typically to maximize the cumulative reward (or return) by choosing a sequence of actions in response to observed states. Conventional RL methods optimize policies by iteratively exploring actions within the environment, gradually learning from rewards. Unlike conventional RL methods, which often learn optimal policies through trial and error, DTs operate on a large offline dataset of pre-collected experiences. This shift to offline data allows DTs to train without requiring real-time environment feedback, which is otherwise costly or even impractical in many applications. 

\begin{figure}[h]
    \centering
    \includegraphics[width=0.45\textwidth]{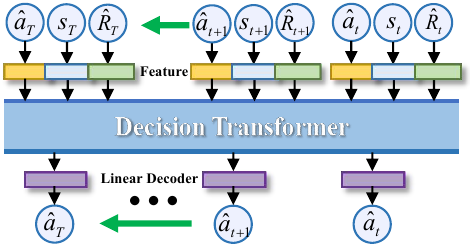}
    \caption{Illustration of decision Transformer structure.}
	\label{DT structure}
\end{figure}

As illustrated in Fig.~\ref{DT structure}, DTs employ the Transformer model, originally designed for sequential data in NLP. In the DT framework, sequences of data samples (comprising states $s_{t}$, actions $\hat{a}_{t}$, and return-to-go $\hat{R}_{t}$) are fed into the Transformer. Return-to-go refers to the cumulative future reward expected from a given state onward, serving as an indicator of the goal or desired outcome. This concept aligns well with RL objectives, as maximizing the return from each state-action pair is key. Within the DT model, causal self-attention mechanisms ensure that each prediction for a future action only considers information from previous or current states and returns, preventing any leakage of future information. 

This autoregressive setup allows the DT model to learn the temporal dependencies between states, actions, and rewards, leveraging patterns within the data to anticipate future actions. Conditioned on both the desired return-to-go and the current state, the model predicts the next action in the sequence. The DT is trained through supervised learning, where it minimizes the loss between its predicted actions and the actual actions observed in the offline dataset. DTs optimize their action predictions by learning directly from pre-recorded trajectories, adjusting the model parameters to replicate the behavior seen in successful sequences. Once trained, DTs can be deployed to make online decisions and generate actions that approximate an optimal policy. One of the primary advantages of DTs is their ability to leverage large and diverse datasets, making them suitable for applications where online exploration is resource-intensive or risky. For instance, in wireless communication networks, where network conditions and traffic patterns evolve dynamically, training a DT on historical data can yield a model capable of optimizing network parameters in real-time without extra exploration.

In summary, DMs and DTs are powerful tools for dealing with complex data and decision-making problems. DMs can generate high-quality data sequence by simulating the noising and denoising processes of data, while DTs can handle decision-making problems through sequence modeling. The combination of these two methods offers a new approach to solving complex decision-making problems.

\section{System Model}
\subsection{Communication Model}
We consider a downlink communication system assisted by an RIS, where the BS is equipped with $M$ antennas, the RIS consists of $N$ elements, and the end user has a single antenna. The channel is assumed to be quasi-static and frequency-flat within a time-slotted system, where each time slot $t\in\left[1,\dots,T\right]$ with a duration of $\tau$. Additionally, the direct link between the BS and the user is assumed to be obstructed due to environmental factors such as buildings. Such conditions often lead to compromised line-of-sight, making it essential to employ an RIS to enhance signal quality.

For a given time slot $t$, the channel coefficient matrix from the BS to the RIS is represented by $\boldsymbol{\mathrm{G}}_{t} \in \mathbb{C}^{N \times M}$ while the channel vector from the RIS to the user is $\boldsymbol{\mathrm{h}}_{t} \in \mathbb{C}^{N \times 1}$. Consequently, the received signal at the user is modeled by
\begin{equation}\label{received signal form 1}
y_{t} = \left( \boldsymbol{\mathrm{h}}_{t}^H \boldsymbol{\Phi}_{t} \boldsymbol{\mathrm{G}}_{t} \right) \boldsymbol{\mathrm{f}}_{t} s_{t} + n_{t},
\end{equation}
where $\boldsymbol{\Phi}_{t} = \mathrm{diag}(\boldsymbol{\phi}_{t})$ represents the beamforming matrix at the RIS and $\boldsymbol{\phi}_{t} = [e^{j\phi_{t,1}}, \dots, e^{j\phi_{t,N}}]^T$ is the beamforming vector. Here, $\phi_{t,n}$ (ranging from $0$ to $2\pi$) refers to the phase reflection coefficients of the $n$-th element and the amplitude coefficients is assumed to be $1$ for maximizing the intensity of reflecting signals. The beamforming vector at the BS, $\boldsymbol{\mathrm{f}}_{t} \in \mathbb{C}^{M \times 1}$, adheres to the power constraint $||\boldsymbol{\mathrm{f}}_{t}||^2 \leq P$. The transmitted symbol,~$s_{t}$, is normalized such that $\mathbb{E}[|s_{t}|^2] = 1$, and $n_{t}$ stands for the additive white Gaussian noise with zero mean and variance $\sigma^2$. Accordingly, the received signal in (\ref{received signal form 1})
can be rewritten as
\begin{equation}
y_{t} = \left( \boldsymbol{\phi}_{t}^T \mathrm{diag}(\boldsymbol{\mathrm{h}}_{t}^H) \boldsymbol{\mathrm{G}}_{t} \right) \boldsymbol{\mathrm{f}}_{t} s_{t} + n_{t},
\end{equation}
where $\boldsymbol{\mathrm{H}}_{t} = \mathrm{diag}(\boldsymbol{\mathrm{h}}_{t}^H) \boldsymbol{\mathrm{G}}_{t}$ represents the cascaded BS-RIS-user channel matrix. Without loss of optimality, we adopt the maximal-ratio transmission (MRT)~\cite{Max_Ratio_Trans_MRT} as the BS beamforming policy for a given $\boldsymbol{\phi}_{t}$ and $\boldsymbol{\mathrm{H}}_{t}$, i.e., $\boldsymbol{\mathrm{f}}^{*}_{t}=\sqrt{P}(\boldsymbol{\phi}_{t}^T \boldsymbol{\mathrm{H}}_{t})^{H}/||\boldsymbol{\phi}_{t}^T \boldsymbol{\mathrm{H}}_{t}||$. 

Our primary goal is to maximize the long-term cumulative data rate over multiple time slots, which can be formulated as
\begin{equation}\label{optimization problem}
\begin{aligned}
\textbf{P}: & \max_{\left\{\boldsymbol{\phi}_{t}\right\}} \sum_{t=1}^{T} \log_{2}\left(1 + \frac{P \left\|\boldsymbol{\phi}_{t}^T \boldsymbol{\mathrm{H}}_{t}\right\|^2}{\sigma^2}\right),\\
& \text{s.t.} \quad |\boldsymbol{\phi}_{t}^{n}| = 1, \forall n \in [1, \dots, N].
\end{aligned}
\end{equation}

\subsection{Channel Correlation Model}
We consider an RIS composed of a rectangular array with $N = N_{1} N_{2}$ elements, where $N_1$ and $N_2$ represent the numbers of elements per column and row, respectively. Each element has a size defined by $\sigma = d_1 d_2$, where $d_1$ and $d_2$ are the numbers of rows and columns, respectively\footnote{The RIS construction method presented in this paper is designed to simplify the derivation of subsequent spatial correlations. Although the correlations associated with RIS may encompass a variety of scenarios, the methodology proposed in this paper is broadly applicable.}. Assuming the RIS is located in the yz-plane, the position of the element at the $n_1$-th row and $n_2$-th column ($n_1\in[1, N_1]$, $n_2\in[1, N_2]$) in three-dimensional space can be expressed as 
\begin{equation}
    \boldsymbol{\mathrm{p}}(n_1,n_2) = [0, (n_1-1)d_1, (n_2-1)d_2]^{T}.
\end{equation}

In an isotropic scattering environment, plane waves with wavelength $\lambda$ impinge on the RIS uniformly from all directions. The azimuth and elevation angles are denoted as $\alpha$ and $\beta$, respectively. The probability density function (PDF) can be represented as~\cite{Rayleigh_Fading_Modeling}
\begin{equation}
    f(\alpha,\beta)=\frac{\cos{(\beta})}{2\pi}, \alpha\in\left[-\frac{\pi}{2},\frac{\pi}{2}\right],\beta\in\left[-\frac{\pi}{2},\frac{\pi}{2}\right].
\end{equation}
The array response vector can be written as~\cite{Rayleigh_Fading_Modeling_func}
\begin{equation}\label{array response vector}
    \boldsymbol{\gamma}(\alpha,\beta)=\left[e^{j\boldsymbol{\mathrm{q}}(\alpha,\beta)^{T}\boldsymbol{\mathrm{p}}(0,0)}, \dots, e^{j\boldsymbol{\mathrm{q}}(\alpha,\beta)^{T}\boldsymbol{\mathrm{p}}(N_1,N_2)}\right]^{T},
\end{equation}
where the wave vector is given by,
\begin{equation}\label{wave vector}
    \boldsymbol{\mathrm{q}}(\alpha,\beta)=\frac{2\pi}{\lambda}\left[\cos{(\alpha)}\cos{(\beta)},\sin{(\alpha)\cos{(\beta)},\sin{(\beta)}}\right]^{T}.
\end{equation}

Further, we analyze channel $\boldsymbol{\mathrm{G}}_{t}=[\boldsymbol{\mathrm{g}}_{t,1},\dots,\boldsymbol{\mathrm{g}}_{t,M}]$ from the BS to the RIS, wherein $\boldsymbol{\mathrm{g}}_{t,m}$ represents the channel from the $m$-th antenna of the BS to the RIS. This channel is composed of $K$ multi-path components, i.e.,
\begin{equation}
    \boldsymbol{\mathrm{g}}_{t,m} = \sum_{k=1}^{K}\frac{\kappa_{m,k}}{\sqrt{K}}\boldsymbol{\gamma}(\alpha_{k},\beta_{k}), \forall t, m\in[1,\dots,M],
\end{equation}
where $\kappa_{m,k}$ denotes the signal attenuation of the $k$-th component with zero mean and $\sigma \mu_{m}$ variance. Moreover, $\mu_{m}$ is the average intensity attenuation. As $K$ approaches infinity, according to the central limit theorem (CLT), the channel can be approximated as the following distribution
\begin{equation}
    \boldsymbol{\mathrm{g}}_{t,m}\xrightarrow{K\rightarrow\infty}\mathcal{CN}(\boldsymbol{0},\sigma \mu_{m}\boldsymbol{\mathrm{R}}), \forall t,
\end{equation}
where $\boldsymbol{\mathrm{R}}$ is the normalized spatial correlation matrix, i.e.,
\begin{equation}\label{spatial correlation matrix 1}
    \boldsymbol{\mathrm{R}}=\mathbb{E}\left[\boldsymbol{\gamma}(\alpha,\beta)\boldsymbol{\gamma}(\alpha,\beta)^{H}\right].
\end{equation}
According to (\ref{array response vector}), the element in the $i$-th row ($i=n_1N_2+n_2$) and the $l$-th column ($l=m_1N_2+m_2$) of the correlation matrix can be expressed as
\begin{equation}\label{spatial correlation matrix 2}
    \boldsymbol{\mathrm{R}}^{i,l}=\mathbb{E}\left[e^{j\boldsymbol{\mathrm{q}}(\alpha,\beta)^{T}(\boldsymbol{\mathrm{p}}(n_1,n_2)-\boldsymbol{\mathrm{p}}(m_1,m_2))}\right].
\end{equation}

\newtheorem{proposition}{Proposition}
\begin{proposition} \label{proposition1}
In the isotropic scattering environment, the elements of the spatial correlation matrix $\boldsymbol{\mathrm{R}}^{i,l}$ are characterized as follows
\begin{equation}\label{spatial correlation matrix sinc}
    \boldsymbol{\mathrm{R}}^{i,l}=\mathrm{sinc}\left(\frac{2\pi d_2}{\lambda}(n_2-m_2)\right),
\end{equation}
where $\mathrm{sinc}(x)=\sin{(x)}/x$.
\end{proposition}
\begin{IEEEproof}
See Appendix \ref{proof of lemma 1}.
\end{IEEEproof}

Eqn. (\ref{spatial correlation matrix sinc}) describes the spatial correlation between the channel elements from the BS to the RIS. This correlation depends on both the distance between the RIS elements and their size. Similarly, the channel elements from the RIS to the user exhibit similar spatial correlations, assuming identical propagation conditions but with a different average intensity attenuation $\mu_{0}$, i.e., $\boldsymbol{\mathrm{h}}_{t}\sim\mathcal{CN}(\boldsymbol{\mathrm{0}},\sigma\mu_{0}\boldsymbol{\mathrm{R}})$. Based on the aforementioned derivation, we can further demonstrate that the cascaded channel also exhibits spatial correlation.

\newtheorem{remark}{Remark}
\begin{remark}\label{remark1}
For the arbitrary column vector of the cascaded channel, $\boldsymbol{\mathrm{h}}^{c}_{t,m}=\mathrm{diag}(\boldsymbol{\mathrm{h}}_{t})\boldsymbol{\mathrm{g}}_{t,m}$, where each element is the product of two complex Gaussian random variables. When $N$ is sufficiently large, according to the CLT, this vector retains spatial correlation among the elements.
\end{remark}

\section{Proposed Solution}
This section introduces a novel approach to RIS-assisted wireless communication systems by integrating DM with DT. The synergy between DM and DT provides a robust framework for inferring full channel state information (FCSI) from partial channel state information (PCSI), thereby optimizing the beamforming strategies in stochastic environments.

\subsection{Diffusion Model Based Channel Acquisition}
In large-scale RIS scenarios with a large number elements, performing channel estimation for each element results in high pilot overhead, which ultimately reduces the available time resources for data transmission and thereby degrading the system rate. To effectively address this issue, we utilize the partial information to infer the global CSI. Specifically, we estimate the CSI for a subset of elements, which inherently reduces pilot overhead, and exploit the spatial correlation between channels to predict the CSI for all elements.

The introduction of DM further enhances our approach. DMs are a class of generative models that iteratively transform random noise into high-quality samples through a two-step process, i.e., the forward diffusion process and the reverse denoising process. In the forward diffusion process, the original data are gradually corrupted by adding noise over a series of steps, ultimately approximating a Gaussian distribution. The reverse denoising process reconstructs the original data by progressively removing the noise using learned denoising functions. By employing DMs, we can effectively infer the CSI for all elements based on the estimated subset, thereby enhancing the accuracy and efficiency of channel estimation.

We first define the estimated cascade channel corresponding to a subset of elements as $\boldsymbol{\hat{\mathrm{H}}}^{\mathrm{p}}_{t}\in\mathbb{C}^{N^{\mathrm{p}}\times M}$, where $N^{\mathrm{p}}$ represents the number of elements for which CSI is estimated. The CSI for all elements is denoted as $\boldsymbol{\hat{\mathrm{H}}}_{t}\in\mathbb{C}^{N\times M}$. The mask ratio is defined as $\rho=1-N^{\mathrm{p}}/N$, indicating the proportion of elements for which the CSI is not directly estimated. Since the CSI is in the complex domain, it is converted into real-valued vectors by vectorizing the real and imaginary parts of the CSI. The resulting vector corresponding to $\boldsymbol{\hat{\mathrm{H}}}_{t}$ is denoted as $\boldsymbol{x}_{t}\in\mathbb{R}^{2NM\times 1}$, while the vector corresponding to $\boldsymbol{\hat{\mathrm{H}}}^{\mathrm{p}}_{t}$ is represented as $\boldsymbol{x}^{\mathrm{p}}_{t}\in\mathbb{R}^{2N^{\mathrm{p}}M\times 1}$. Then, we define the forward diffusion process, which involves progressively adding noise to the original channel data, i.e.,
\begin{equation}
    \boldsymbol{x}^{(k)}_{t}=\sqrt{1-b_{k}}\boldsymbol{x}^{(k-1)}_{t}+\sqrt{b_{k}}\boldsymbol{\epsilon}^{(k-1)}_{t}, k=1,\dots,K,
\end{equation}
where $\boldsymbol{x}^{(k)}_{t}$ is the latent data with noise at diffusion step $k$, $b_{k}$ denotes variance of the added noise at step $k$, and $\boldsymbol{\epsilon}^{k-1}_{t}$ is Gaussian noise that follows the standard normal distribution. Notice that $\boldsymbol{x}_{t}^{(0)}$ corresponds to original data $\boldsymbol{x}_{t}$, then the noise sample at step $k$ can be represented as 
\begin{equation}
    \boldsymbol{x}^{(k)}_{t}=\sqrt{\bar{a}_{k}}\boldsymbol{x}^{(0)}_{t}+\sqrt{1-\bar{a}_{k}}\epsilon^{(0)}_{t}, k=1,\dots,K,
\end{equation}
where $\bar{a}_{k}=\prod_{i=1}^{k}(1-b_{i})$. As each diffusion step is added by a Gaussian noise, we have  distributions
\begin{equation}\label{forward diffusion distribution}
\begin{aligned}
    q(\boldsymbol{x}^{(k)}_{t}|\boldsymbol{x}^{(k-1)}_{t})&\sim\mathcal{N}(\boldsymbol{x}^{(k)}_{t};\sqrt{1-b_{k}}\boldsymbol{x}^{(k-1)}_{t},b_{k}\boldsymbol{\mathrm{I}}),\\
    q(\boldsymbol{x}^{(k)}_{t}|\boldsymbol{x}^{(0)}_{t})&\sim\mathcal{N}(\boldsymbol{x}^{(k)}_{t};\sqrt{\bar{a}_{k}}\boldsymbol{x}^{(k-1)}_{t},(1-\bar{a}_{k})\boldsymbol{\mathrm{I}}).
\end{aligned}
\end{equation}
As $K$ increases, $\bar{a}_{k}$ approaches $0$, and the probability distribution of $\boldsymbol{x}^{(K)}_{t}$ converges towards a Gaussian distribution. 

During the training process of the network, we aim to predict the noise artificially added during the forward diffusion process and progressively remove it from $\boldsymbol{x}^{(K)}_{t}$
in order to recover original channel data $\boldsymbol{x}^{(0)}_{t}$. This approach enables us to generate the channel from arbitrary Gaussian noises. However, such generation with the classical DM is inherently random, which can lead to an inability to accurately generate CSI. 

To address this limitation, we propose a diffusion generation method based on partial information, wherein a subset of channel data 
$\boldsymbol{x}_{t}^{\mathrm{p}}$ is utilized as conditional information. This allows the DM to learn the latent spatial relationship between the partial channel and the full channel, thereby guiding the generation of real-time CSI. Specifically, in the reverse denoising phase, if we can get the probability distribution $q(\boldsymbol{x}^{(k-1)}_{t}|\boldsymbol{x}^{(k)}_{t})$ we can derive 
$\boldsymbol{x}^{(0)}_{t}$ starting from $\boldsymbol{x}^{(K)}_{t}$ step by step. However, computing $q(\boldsymbol{x}^{(k-1)}_{t}|\boldsymbol{x}^{(k)}_{t})$ is quite challenging, as it requires the joint probability distribution of  data across all the steps. 

To approximate this distribution, we employ a neural network model $q_{\theta}(\boldsymbol{x}^{(k-1)}_{t}|\boldsymbol{x}^{(k)}_{t}, \boldsymbol{x}_{t}^{\mathrm{p}})$, where $\theta$ is the network parameters. Furthermore, as indicated in~\cite{DDPM}, the conditional probability distribution, $q(\boldsymbol{x}^{(k-1)}_{t}|\boldsymbol{x}^{(k)}_{t},\boldsymbol{x}^{(0)}_{t})$, becomes tractable when original data  $\boldsymbol{x}^{(0)}_{t}$ is known. This is because the data at any step can be represented by the original data, with the mean $\mu_{k}$ and variance $\Sigma_{k}$ of the conditional distribution 
$q(\boldsymbol{x}^{(k-1)}_{t}|\boldsymbol{x}^{(k)}_{t},\boldsymbol{x}^{(0)}_{t})$ as
\begin{equation}
\begin{aligned}
    \mu_{k}&=\frac{1}{\sqrt{1-b_{k}}}\left(\boldsymbol{x}_{t}^{(k)}-\frac{b_{k}}{\sqrt{1-\bar{a}_{k}}}\boldsymbol{\epsilon}_{t}^{(k)}\right),\\
    \Sigma_{k}&=\frac{1-\bar{a}_{k-1}}{1-\bar{a}_{k}}b_{k},
\end{aligned}
\end{equation}
from (\ref{forward diffusion distribution}). To ensure that the function $q_{\theta}(\boldsymbol{x}^{(0)}_{t}|\boldsymbol{x}_{t}^{\mathrm{p}})$ approximates $q(\boldsymbol{x}^{(0)}_{t})$, it is common to minimize the cross entropy (CE) between the two distributions, i.e.,
\begin{equation}\label{cross_entropy}
    L_{\mathrm{CE}}=-\mathbb{E}_{q(\boldsymbol{x}^{(0)}_{t})}\log\left(q_{\theta}(\boldsymbol{x}^{(0)}_{t}|\boldsymbol{x}_{t}^{\mathrm{p}})\right).
\end{equation}
This objective can be further formulated as minimizing a combination of multiple terms involving Kullback-Leibler (KL) divergence and entropy, i.e.,
\begin{equation}
\begin{aligned}
    L=&\mathbb{E}_{q}\left[D_{\mathrm{KL}}\left(q(\boldsymbol{x}^{(K)}_{t}|\boldsymbol{x}^{(0)}_{t})||q_{\theta}(\boldsymbol{x}_{t}^{(K)}|\boldsymbol{x}_{t}^{\mathrm{p}})\right)\right.\\
    &\left.+\sum_{k=2}^{K}D_{\mathrm{KL}}\left(q(\boldsymbol{x}^{(k-1)}_{t}|\boldsymbol{x}^{(k)}_{t},\boldsymbol{x}^{(0)}_{t})||q_{\theta}(\boldsymbol{x}_{t}^{(k-1)}|\boldsymbol{x}_{t}^{(k)},\boldsymbol{x}_{t}^{\mathrm{p}})\right)\right.\\
    &\left.-\log q_{\theta}(\boldsymbol{x}_{t}^{(0)}|\boldsymbol{x}_{t}^{(1)},\boldsymbol{x}_{t}^{\mathrm{p}})\right].
\end{aligned}
\end{equation}

\begin{proposition} \label{proposition2}
Since the output of neural network $q_{\theta}(\cdot)$ represents the mean and variance of a multivariate Gaussian distribution, the loss function can be further simplified as
\begin{equation}\label{DM loss function simple}
\begin{aligned}
    &L=L_{0}+\sum_{k=1}^{K-1}L_{1,k}+L_{2},
\end{aligned}
\end{equation}
where 
\begin{equation}\label{DM loss function expand}
\begin{aligned}
    &L_{0}=D_{\mathrm{KL}}\left(q(\boldsymbol{x}^{(K)}_{t}|\boldsymbol{x}^{(0)}_{t})||q_{\theta}(\boldsymbol{x}_{t}^{(K)}|\boldsymbol{x}_{t}^{\mathrm{p}})\right),\\
    &L_{1,k}=\mathbb{E}\left[||\boldsymbol{\epsilon}_{t}^{(k)}-\tilde{\boldsymbol{\epsilon}}_{\theta,t}^{(k)}(\boldsymbol{x}_{t}^{(k)},\boldsymbol{x}_{t}^{\mathrm{p}},k)||^{2}\right],\\
    &L_{2}=-\log q_{\theta}(\boldsymbol{x}_{t}^{(0)}|\boldsymbol{x}_{t}^{(1)},\boldsymbol{x}_{t}^{\mathrm{p}}).
\end{aligned}
\end{equation}
\end{proposition}
\begin{IEEEproof}
See Appendix \ref{proof of diffusion loss function}.
\end{IEEEproof}
Notably, both $L_{0}$ and $ L_{2}$ can be neglected during the network training since $\boldsymbol{x}_{t}^{K}$ is a artificially Gaussian noise and $\boldsymbol{x}_{t}^{(0)}$ is the original data. The network outputs the predicted artificially noise, denoted as $\tilde{\boldsymbol{\epsilon}}_{\theta,t}^{(k)}(\boldsymbol{x}_{t}^{(k)},\boldsymbol{x}_{t}^{\mathrm{p}},k)$.

By minimizing this loss function, we reduce the discrepancy between the network's predictive noise and the artificially added noise during the forward diffusion process. Consequently, even with partial channel information, we can gradually eliminate the initially added Gaussian noise, ultimately achieving data imputation and recovering real-time CSI for all RIS elements.


\subsection{Policy Generalization with Decision Transformer}
Solving the optimization problem 
$\boldsymbol{\mathrm{P}}$ in (\ref{optimization problem}) is relatively straightforward. Once the CSI is obtained, conventional optimization methods, such as SDR and RL, can be applied to obtain the high-quality beamforming policies. However, these methods typically assume perfect knowledge of the CSI or a static CSI environment. When faced with previously unseen scenarios, the performance of well-trained beamforming policies based on RL models may degrade, demanding to retrain the models from the beginning using newly collected data .

\begin{figure*}[htbp]
    \centering
    \includegraphics[width=1.0\textwidth]{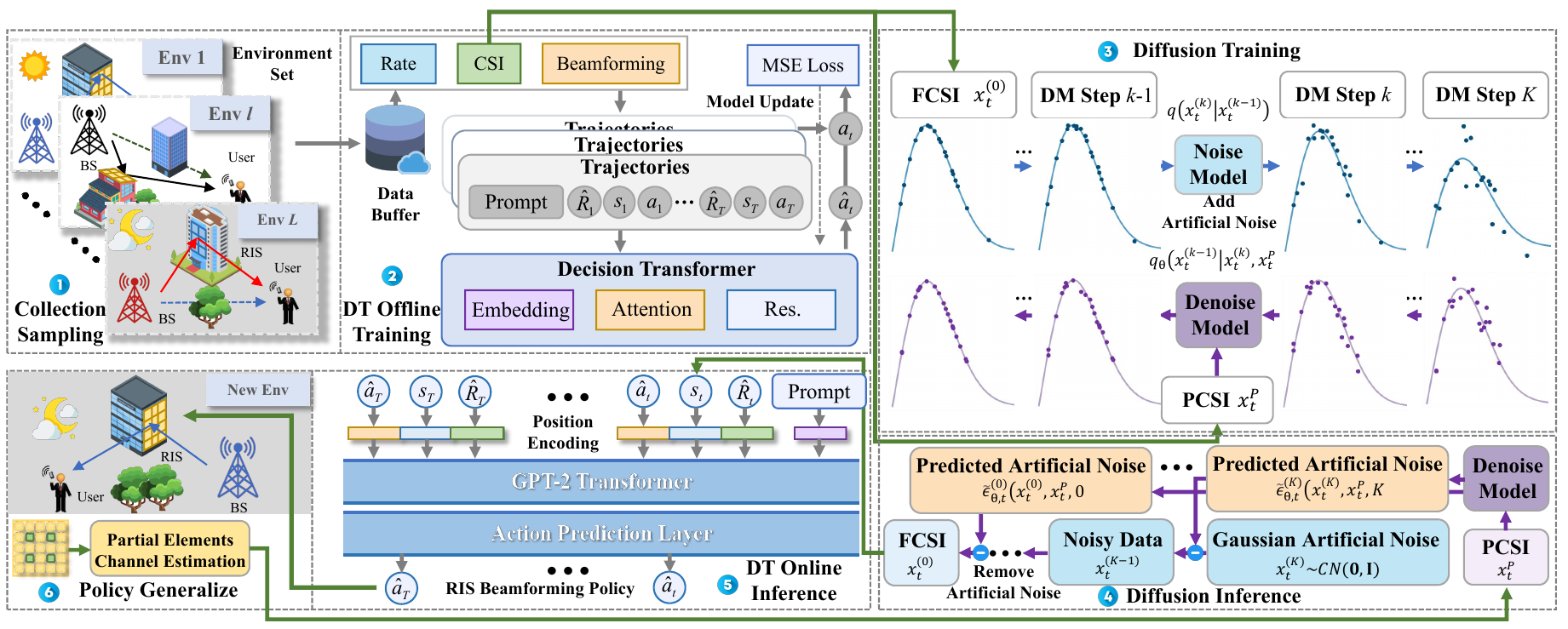}
    \caption{The proposed DEDT framework. It integrates a DM to derive FCSI from PCSI by leveraging spatial correlations. The DT component utilizes this channel information to rapidly generate efficient RIS beamforming policies. The framework facilitates online fine-tuning and quick adaptation in new scenarios, significantly enhancing beamforming efficacy and system adaptability.}
	\label{whole structure}
\end{figure*}

To solve this issue, we introduce DT to make beamforming decisions with stronger generalization in new scenarios relative to RL. Recent advancements in GPT have demonstrated that Transformer architectures are highly effective in decision-making tasks. Unlike traditional RL approaches, which require extensive online interactions with the environment, DT operates on offline datasets. It reframes the optimization of value or policy functions from RL into a self-regressive prediction of actions based on the sequences of historical states, actions, and rewards. Consequently, the first step is to map the elements of the optimization problem into the fundamental RL tuples, consisting of states, actions and rewards, as detailed below.

\emph{State}: The state, denoted as $\boldsymbol{s}_{t}$, including the cascade CSI between the BS, RIS, and the user, generated by the DM.

\emph{Action}: Upon obtaining the state $\boldsymbol{s}_{t}$, a corresponding RIS beamforming policy to be designed, which is represented as action $\boldsymbol{a}_{t}$.

\emph{Reward}: Defined as the instantaneous data rate received by the user at time slot $t$, denoted by $r_{t}=\log_{2}(1+P \left\|\boldsymbol{\phi}_{t}^T \boldsymbol{\mathrm{H}}_{t}\right\|^2 / \sigma^2)$.

The corresponding objective function can be expressed as the maximization of the cumulative expected reward, i.e., $\mathbb{E}\left[\sum_{t=1}^{T}r_{t}\right]$. We further define the sample sequence for DT training as 
\begin{equation}
    \tau_{t} = \left(\hat{R}_{1},s_{1},a_{1},\dots,\hat{R}_{t},s_{t},a_{t}\right).
\end{equation}
It is also known as the trajectory, which encompasses the states, actions, and rewards over the entire time horizon $T$ in an episode. Here, $\hat{R}_{t}=\sum_{t'=t}^{T}r_{t'}$ represents future cumulative rewards, indicating the sum of rewards from time slot $t$ to the final time slot $T$. This enables the DT to generate actions based on anticipated future rewards, rather than solely relying on historical rewards.

The DT uses the entire sequence of accumulated rewards, states, and actions to directly model and predict the next optimal action. During the training phase, the DT is provided with trajectories $\tau_{t}$. Each sample can be collected offline from other optimization methods. Specifically, action samples are generated via near-optimal policies obtained from convergent RL algorithms or other heuristic algorithms. By leveraging a large offline dataset of such trajectories, the DT is trained in a supervised manner to minimize the difference between predicted actions and the actual near-optimal actions, hence optimizing beamforming decisions across diverse scenarios.

\section{Diffusion-Enhanced Decision Transformer}
In this subsection, we will discuss the entire framework of our proposed DEDT method, as depicted in Fig.~\ref{whole structure}.

\subsection{Training Process for Channel Acquisition}
\begin{figure}[h]
    \centering
    \includegraphics[width=0.45\textwidth]{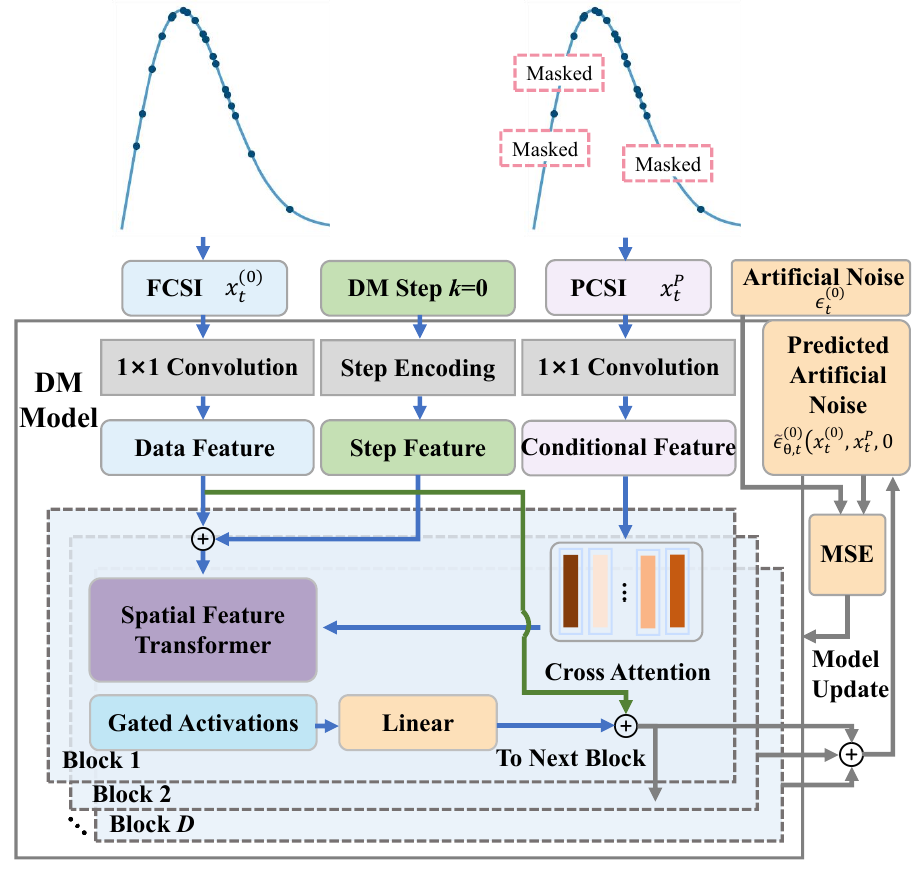}
    \caption{The DM training process. It predicts artificial noise using conditional and full CSI by integrating convolutional networks and spatial feature Transformers to extract features. The model is updated by calculating MSE with actual added artificial noise.}
	\label{DM Training}
\end{figure}

To obtain CSI before decision on beamforming, we initially focus on the corresponding DM part. The whole process of DM is made up of the training and inference parts. As shown in Fig.~\ref{DM Training}, the training phase of the designed DM enables the model to extrapolate FCSI from PCSI enhanced by a spatial feature Transformer. It aims at capturing spatial relationship and correlation among RIS elements, which are essential for the accuracy of the model's output.

Initially, both the FCSI after the $k$-th diffusion step and the PCSI are processed through input embedding and conditional feature embedding layers, mapping high-dimensional vectors to lower-dimensional feature spaces. This facilitates the decoupling of complex features, transforming irregular, nonlinear characteristics into formats more amenable to neural processing. Subsequently, diffusion step $k$ is mapped to the same dimensional features through a step embedding layer.

Then the feature vector corresponding to the PCSI needs to undergo cross-attention computation with the full state feature vector to analyze their correlations. The computed cross-attention and feature vectors are then fed into a spatial feature Transformer. This Transformer leverages the attention mechanism to analyze the importance of certain elements in relation to the known partial elements, quantifying this importance with attention scores while suppressing irrelevant features. The spatial feature Transformer operates by capturing the complex spatial dependencies among RIS elements. It applies multi-head self-attention mechanisms to weight the contribution of each RIS element, which enhances the model's ability to make accurate predictions about the unknown parts of the channel state. It provides a comprehensive understanding of the full channel state estimates from the partial channel.

In the final stages of the training phase, the model employs a combination of convolutional, residual, and linear transformations to refine predictive noise. The convolutional layers are particularly effective at extracting local dependencies between neighboring RIS elements, which is crucial to capture fine-grained CSI. The residual connections help mitigate potential gradient vanishing issues in deep networks, while linear transformations integrate features from multiple layers, improving the model's overall representation capacity. This predictive noise is crucial to reconstruct the channel state during the inference phase. By iterating through these transformations, the model becomes capable of reliably predicting channel states under various conditions, thereby providing a strong foundation for subsequent beamforming optimization. The network's final output of predictive noise is compared to the actual noise added at step $k$, and the parameters of the entire network are updated through gradient backpropagation based on the minimization of the loss function in~(\ref{DM loss function simple}).

\begin{algorithm}
\caption{The DEDT training process for obtaining channel}\label{DEDT training pseudo code 1}
\KwIn{Randomly initialize the network parameters $\theta_{0}$ for DM, the learning rate $\lambda_{1}$, the diffusion step $K$, the noise's variance $b_{k},\forall k$,  the iteration numbers $I_{1}$, and the maximum time slots $T$.}
\KwOut{The generated channel $\hat{\boldsymbol{H}}_{t}$.}

\For{$\mathrm{each}$ $\mathrm{iteration}$ $i = 1,2,\dots,I_{1}$}{
    Sample the cascaded channel $\hat{\boldsymbol{\mathrm{H}}}_{t}$ and randomly mask the elements to get $\hat{\boldsymbol{\mathrm{H}}}_{t}^{\mathrm{p}}$.\\
    Transform the channel $\boldsymbol{\mathrm{H}}_{t}$ and $\hat{\boldsymbol{\mathrm{H}}}_{t}^{\mathrm{p}}$ into vector $\boldsymbol{x}_{t}^{(0)}$ and $\boldsymbol{x}_{t}^{\mathrm{p}}$.\\
    \For{$k = 1,2,\dots,K$}{
        Generate the artificially Gaussian noise $\boldsymbol{\epsilon}_{t}^{(k)}$.\\
        Get the noisy data $\boldsymbol{x}^{(k)}_{t}=\sqrt{1-b_{k}}\boldsymbol{x}^{(k-1)}_{t}+\sqrt{b_{k}}\boldsymbol{\epsilon}^{(k-1)}_{t}$.\\
        Compute the loss function $L_{1,k}=\mathbb{E}\left[||\boldsymbol{\epsilon}_{t}^{(k)}-\tilde{\boldsymbol{\epsilon}}_{\theta,t}^{(k)}(\boldsymbol{x}_{t}^{(k)},\boldsymbol{x}_{t}^{\mathrm{p}},k)||^{2}\right]$.\\
    }
    Update the parameter $\theta_{i+1}\leftarrow \theta_{i}-\lambda_{2}\sum_{k=2}^{K}\nabla_{\theta}L_{1,k}$.
}

\end{algorithm}

The training process for obtaining channel is summarized in \textbf{Algorithm~\ref{DEDT training pseudo code 1}}. It begins with initializing network parameters for the DM and key variables, such as the learning rate, diffusion steps, noise variance, iteration numbers, and maximum time slots. In the training algorithm, we first sample a real channel and apply a mask to simulate the partial CSI, followed by noise generation in each diffusion step. The loss function is then computed by comparing the predicted noise with the true noise at each step. The gradients are computed and used to update the model parameters. The process involves of a series of steps,  beginning with sampling the cascaded channel, applying a mask, and transforming it into vectors (see lines $1$-$3$). During each diffusion step, artificially Gaussian noise is generated, and noisy data is calculated. The loss function is computed based on the difference between the generated noise and the predicted noise by the model (see lines $4$-$7$). Finally, the model parameters are updated using the computed gradients and learning rate (see line $8$).

\subsection{Training Process for RIS Beamforming}
Building on the diffusion model framework for channel acquisition, we further integrate a DT to efficiently optimize RIS beamforming strategies. Initially, offline sequence data are collected from various scenarios (Env $1$, $\cdots$, Env $L$) for the pre-training of the DT. This includes user-received rates, CSI, and near-optimal RIS beamforming strategies, all stored in a data buffer $\mathcal{B}$. To differentiate between scenarios, unique prompts are designed for each one, incorporating the user's anticipated data rate requirements. These prompts assist the Transformer model to recognize specific scenarios and provide tailored instructions. The prompts are designed to capture not just the temporal dynamics of the system but also the key contextual features such as channel quality, user mobility, and interference, which influence RIS beamforming decisions.

Each offline data is then used as input for the DT. The states, actions, and returns-to-go are tokenized through an embedding layer and concatenated with tokenized temporal encoding information. These concatenated feature vectors are then passed into an attention module, where multi-layer self-attention mechanisms capture correlations and feature dependencies within the sequences. The self-attention mechanism enables the model to learn long-range dependencies between states and actions across multiple time steps, which is crucial for accurately modeling the RIS optimization problem over time. Following this, the data flows through a residual network. The predicted actions $\hat{\boldsymbol{a}}_{t}$ are output through a linear activation layer. The entire prediction process of the network can be represented as $\hat{\boldsymbol{a}}_{t}=f_{\omega}(\tau_{t})$, where $\omega$ represents the parameters of the DT network. The objective is to minimize the discrepancy between the predicted actions $\hat{\boldsymbol{a}}_{t}$ and the near-optimal actions $\boldsymbol{a}_{t}$ from the corresponding dataset, i.e.,
\begin{equation}
    L=\mathbb{E}_{\tau_{t}\in\tilde{\mathcal{B}}}\left[(f_{\omega}(\tau_{t})-\boldsymbol{a}_{t})^{2}\right],
\end{equation}
where $\tilde{\mathcal{B}}$ is a mini-batch sampled from the data buffer $\mathcal{B}$.

\begin{algorithm}
\caption{The DEDT training process for RIS beamforming}\label{DEDT training pseudo code 2}
\KwIn{Randomly initialize the network parameters $\omega_{0}$ for DT, the data buffer $\mathcal{B}$, the learning rate $\lambda_{2}$, the iteration numbers $I_{2}$, and the maximum time slots $T$.}
\KwOut{The optimal RIS beamforming matrix $\boldsymbol{\Phi}_{t}$.}
\For{$\mathrm{Env}$ $n=1,2,\dots,L$}{
Learn a well-trained RL algorithm under Env $n$. Use the corresponding policy to generate trajectory $\tau_{t}, \forall t\in\left[1,\dots,T\right]$ and store it in the data buffer $\mathcal{B}$.
}

\For{$\mathrm{each}$ $\mathrm{iteration}$ $i = 1,2,\dots,I_{2}$}{
    Get a batch set $\tilde{B}$ from the replay buffer $\mathcal{B}$.\\
    \For{$t = 1,2,\dots,T$}{
    Get predicted action $\hat{a}_{t}=f_{\omega}(\tau_{t}), \forall \tau_{t}\in\tilde{B}$.\\
    }
    Compute loss $L = \mathbb{E}_{\tau_{t}\in \tilde{\mathcal{B}}}\left[(f_{\omega}(\tau_{t})-a_{t})^{2}\right]$.\\
    Update the parameter $\omega_{i+1}\leftarrow \omega_{i}-\lambda_{1}\nabla_{\omega}L$.\\
}
\end{algorithm}
The training process for learning the optimal RIS beamforming policy is detailed in Algorithm~\ref{DEDT training pseudo code 2}. It starts by initializing network parameters for the DT, setting up a data buffer and defining the learning rate, iteration numbers, and maximum time slots. The data buffer stores trajectories generated from various scenarios, capturing a diverse set of experiences for the model to learn from. The process includes training a RL algorithm across different environments to generate trajectories, which are subsequently stored in the data buffer (see lines $1$-$2$). During each iteration, a batch set is retrieved from the buffer, and predicted actions are calculated for each time step. The loss is computed by measuring the difference between the predicted and actual actions, after which the network parameters are updated accordingly (see lines $3$-$8$).

\subsection{The DEDT inference process}
Once the diffusion model has been fully trained, we will deploy it into a new environment. Similar to the training phase, we initially require the acquisition of channel state information. Initially, a partial channel estimation $\boldsymbol{x}_{t}^{\mathrm{p}}$ is performed using traditional methods. Starting from an initial random Gaussian noise  $\boldsymbol{x}_{t}^{(0)}$, the model iteratively applies the predicted noise $\tilde{\boldsymbol{\epsilon}}_{\theta,t}^{(k)}(\boldsymbol{x}_{t}^{(k)},\boldsymbol{x}_{t}^{\mathrm{p}},k)$ to gradually reconstruct the channel state representation. This iterative process is carefully designed to refine the noise model at each step, ensuring that each successive approximation is progressively closer to the true channel state, i.e.,
\begin{equation}\label{data denoise}
    \boldsymbol{x}_{t}^{(k-1)}=\frac{1}{\sqrt{1-b_{k}}}\left(\boldsymbol{x}_{t}^{(k)}-\frac{b_{k}}{1-\bar{a}_{t}}\tilde{\boldsymbol{\epsilon}}_{\theta,t}^{(k)}\right) + \frac{1-\bar{a}_{k-1}}{1-\bar{a}_{k}}b_{k}\boldsymbol{\epsilon},
\end{equation}
where $\boldsymbol{\epsilon}$ represents a random Gaussian noise. Here the reparameterization trick is applied during the denoising process to ensure both stability and accuracy, which allows for more precise control of the noise in the reconstructed channel states. The reparameterization trick ensures that the model can sample from a distribution with fixed noise variance while maintaining gradient flow for backpropagation. This enables more efficient learning and smoother inference in real-time applications.

The ability to quickly and accurately predict the channel state from incomplete information makes the DM particularly valuable for real-time decision-making in the new environments. This capability not only enhances the responsiveness of RIS-assisted communication systems but also significantly reduces the pilot overhead required for channel estimation. By leveraging the noise modeling and denoising process, the model can efficiently estimate the channel state without requiring a full pilot transmission, thus reducing overhead and improving the system’s real-time adaptability.

For RIS beamforming, the pre-trained DT model cannot be directly applied because the new environments may introduce unique prompts that differ from the offline training environment. This requires additional fine-tuning of the pre-trained model. Initially, sub-optimal beamforming actions are generated using traditional methods, such as non-converged RL policies, to generate a few-shot sample trajectory sequences for fine-tuning. Subsequently, we freeze the initial layers of the DT's network parameters, only fine-tuning the final linear activation layer. The decision to freeze the initial layers is based on the assumption that lower layers capture more general, transferable features, such as spatial patterns in channel states, that are applicable across environments. 

Meanwhile, the final layers focus on the task-specific adjustments required by new scenarios, such as optimal RIS beamforming actions for different data rates or channel conditions. The fine-tuned DT model can then be deployed for real-time decision-making in the new environments. Specifically, the DM is first used to obtain CSI as state input $\boldsymbol{s}_{t}$. Taking the new environment's required data rate as the prompt, the anticipated cumulative rewards serve as the initial return-to-go $\hat{R}_{t}$. Based on this setup, the DT model continually infers and predicts actions $\hat{\boldsymbol{a}}_{t}$, further interacting with the environment to obtain corresponding data rate rewards and updated CSI. This iterative process generates the action sequence to maximize cumulative rewards.

\begin{algorithm}
\caption{The DEDT inference process}\label{DEDT inference pseudo code}
\KwIn{New $\mathrm{Env}$, few-shot samples $\hat{B}$, diffusion step $K$, the noise's variance $b_{k},\forall k$, maximum time slots $T$, target return-to-go $\hat{R}_{1}$, and initial input sequence $\tau^{\mathrm{in}}_{1}=(\hat{R}_{1})$, the number of RIS elements for channel estimation $N^{\mathrm{p}}$}
\KwOut{The optimal RIS beamforming matrix $\boldsymbol{\Phi}_{t}$.}
Fine-tune the pre-trained DT model with $\hat{B}$.\\
\For{$t=1,2,\dots, T$}{
    Estimate the PCSI $\hat{\boldsymbol{\mathrm{H}}}_{t}$ and convert it to $\boldsymbol{x}_{t}^{\mathrm{p}}$.\\ Generate a random Gaussian noise $\boldsymbol{x}_{t}^{(K)}$.\\
    \For{$k=K,\dots,1$}{
    Predict noise $\tilde{\boldsymbol{\epsilon}}_{\theta,t}^{(k)}(\boldsymbol{x}_{t}^{(k)},\boldsymbol{x}_{t}^{\mathrm{p}},k)$.\\
    Denoise the data $\boldsymbol{x}_{t}^{(k-1)}=\frac{1}{\sqrt{1-b_{k}}}\left(\boldsymbol{x}_{t}^{(k)}-\frac{b_{k}}{1-\bar{a}_{t}}\tilde{\boldsymbol{\epsilon}}_{\theta,t}^{(k)}\right) + \frac{1-\bar{a}_{k-1}}{1-\bar{a}_{k}}b_{k}\boldsymbol{\epsilon}$} 
    Transform the data $\boldsymbol{x}_{t}^{0}$ into channel state $s_{t}$.\\
    Update the trajectory sequence $\tau^{\mathrm{in}}_{t}\leftarrow(\tau^{\mathrm{in}}_{t},s_{t})$.\\
    Get predicted action $\hat{a}_{t}=f_{\omega}(\tau^{\mathrm{in}}_{t})$.\\
    Interact with the environment to get reward $r_{t}$.\\
    Update the return-to-go $\hat{R}_{t+1}=\hat{R}_{t}-r_{t}$.\\
    Update the sequence $\tau^{\mathrm{in}}_{t+1}\leftarrow (\tau^{\mathrm{in}}_{t},\hat{a}_{t},\hat{R}_{t+1}).$
}
\end{algorithm}

This online fine-tuning inference methodology enables our framework to swiftly adapt to the new scenarios, facilitating the rapid transfer of beamforming policies while minimizing channel estimation overhead. As indicated in Algorithm~\ref{DEDT inference pseudo code}, it begins by fine-tuning a pre-trained DT model using few-shot samples. For each time slot, the DM iteratively predicts noise and denoises data in a reverse diffusion process based on the estimated PCSI. The denoised data is transformed into channel state and appended to the input sequence (see lines $2$-$8$). A predicted action is generated based on the updated sequence, and the environment is interacted with to obtain a reward. The return-to-go is updated, and the sequence is then updated for further action prediction in the next time slot (see lines $8$-$13$).

\section{Simulation Results}
To evaluate the proposed DEDT framework in RIS-assisted communication systems, we provide the simulation settings and results. The BS is equipped with a uniform linear array with $M=4$ antennas. The average
intensity attenuation of the channel are all set to $0.5$. We utilize the default parameter settings in~\cite{IRS_DM} as the simulation environment for offline pre-training. Furthermore, we change the channel parameters and BS-RIS distances to simulate different wireless environments.

Then, we introduce the detailed network settings of the proposed DEDT framework. The diffusion training for the DM component is executed over $K=500$ steps. The variation of noise parameter 
$\beta_{k}$ from $10^{-4}$ to $0.02$ ensures that the model can adapt to both subtle and significant variations in the channel state information during the diffusion process. Each block within the DM includes one-dimensional convolutional layers and linear layers. The DT segment of our framework comprises three Transformer blocks. Each block has a hidden layer dimension of $256$ and incorporates a dropout rate of $0.1$ to prevent overfitting during the intensive training process. This configuration is designed to handle the sequential decision-making process effectively, enabling the DT to learn and adapt from historical data and predict optimal actions under new conditions. The entire model is trained on high-performance NVIDIA RTX A6000 GPU hardware, utilizing the AdamW optimizer with a learning rate of $10^{-4}$. 

\begin{figure}[htbp]
    \centering
    \includegraphics[width=0.45\textwidth]{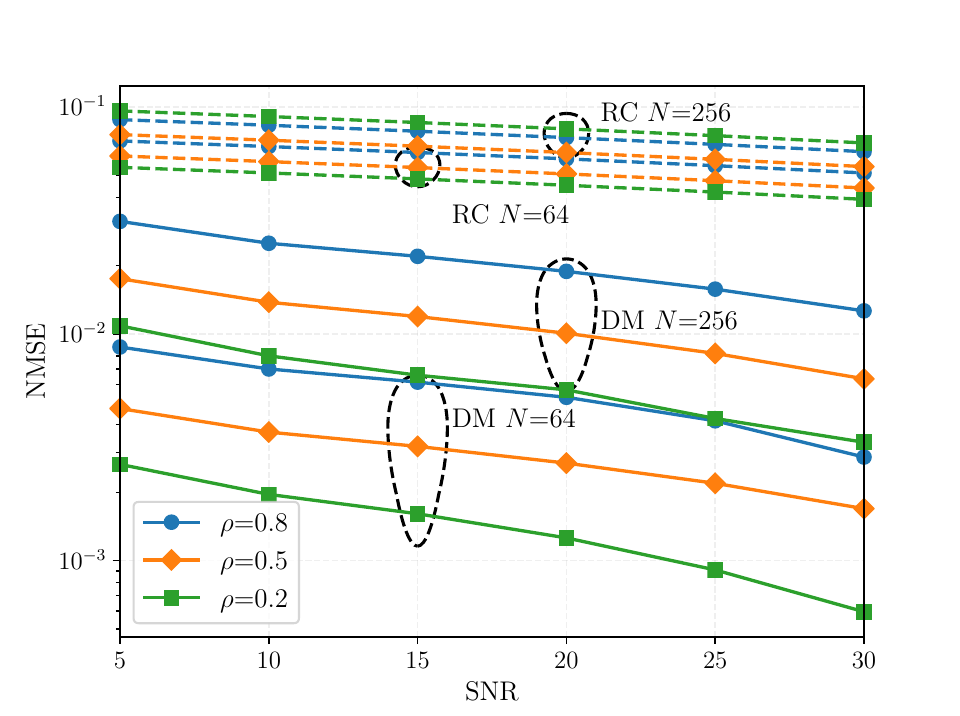}
    \caption{NMSE versus SNR under different channel mask ratio $\rho$ and RIS elements $N$.}
    \label{NMSE}
\end{figure}

\begin{figure}[htbp]
    \centering
    \includegraphics[width=0.45\textwidth]{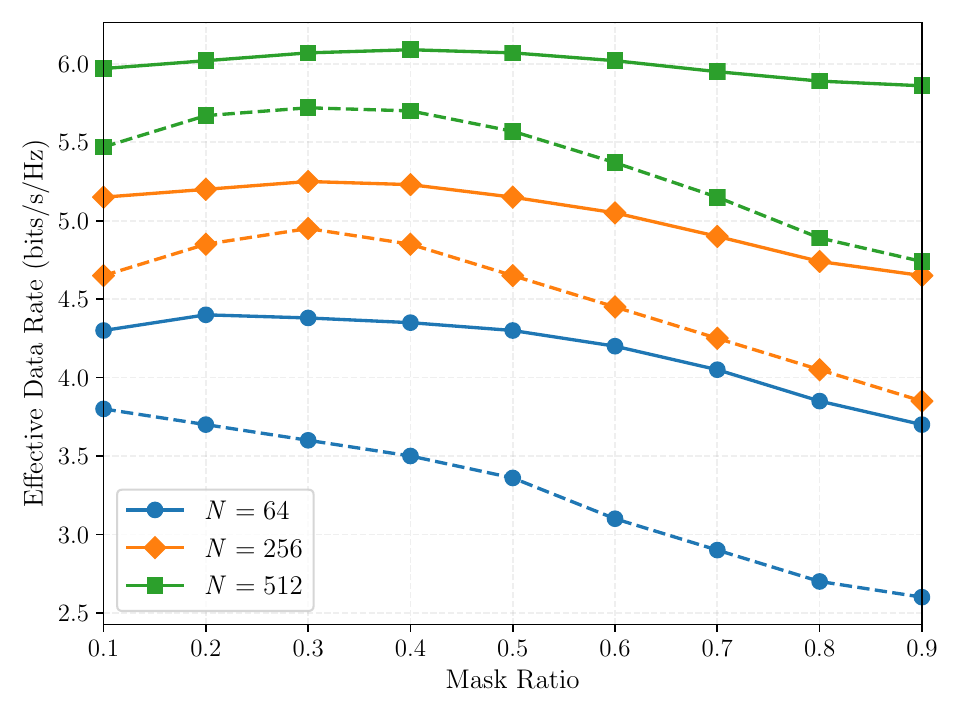}
    \caption{Performance versus channel mask ratio under different RIS elements $N$. Solid line: DEDT, Dashed line: RCDT}
    \label{optimal_mask_ratio}
\end{figure}

We first demonstrate the channel generation capabilities of the proposed conditional information-based DEDT model, as shown in Fig.~\ref{NMSE}, which illustrates the variations in average normalized mean-square error (NMSE) with different mask ratios and RIS element numbers. In order to compare with the proposed channel generation method (DM), we use the random channel (RC) approach, where the remaining channels are randomly generated based on the unmasked RIS elements, as the baseline method. From the figure, the NMSE decreases with increasing signal-to-noise ratio (SNR), which is an expected outcome. Moreover, at a fixed SNR level, the NMSE increases with the mask ratio. This phenomenon can be attributed to the fact that a higher mask ratio results in reduced auxiliary information being fed into the DM network. Consequently, this diminishes the model's ability to effectively discern the relationship between FCSI and PCSI, thereby impacting the ultimate channel generation quality.

Fig.~\ref{optimal_mask_ratio} compares the effective data rate performance of the proposed DEDT algorithm with the baseline RCDT algorithm across varying mask ratios. The effective data rate represents the actual data rate after subtracting the channel estimation overhead associated with the unmasked RIS elements. In the RCDT method, the channels are generated based on the RC approach, followed by decision-making with DT. The results demonstrate that DEDT consistently outperforms RCDT, achieving higher data rates under the same mask ratio. Notably, DEDT exhibits an optimal mask ratio, where the effective data rate reaches its peak. This highlights the algorithm’s ability to effectively balance channel estimation sparsity and spectral efficiency. Beyond this range, the performance of DEDT slightly decreases, reflecting the trade-off between maintaining channel estimation cost and system performance.

\begin{figure}[htbp]
    \centering
    \includegraphics[width=0.45\textwidth]{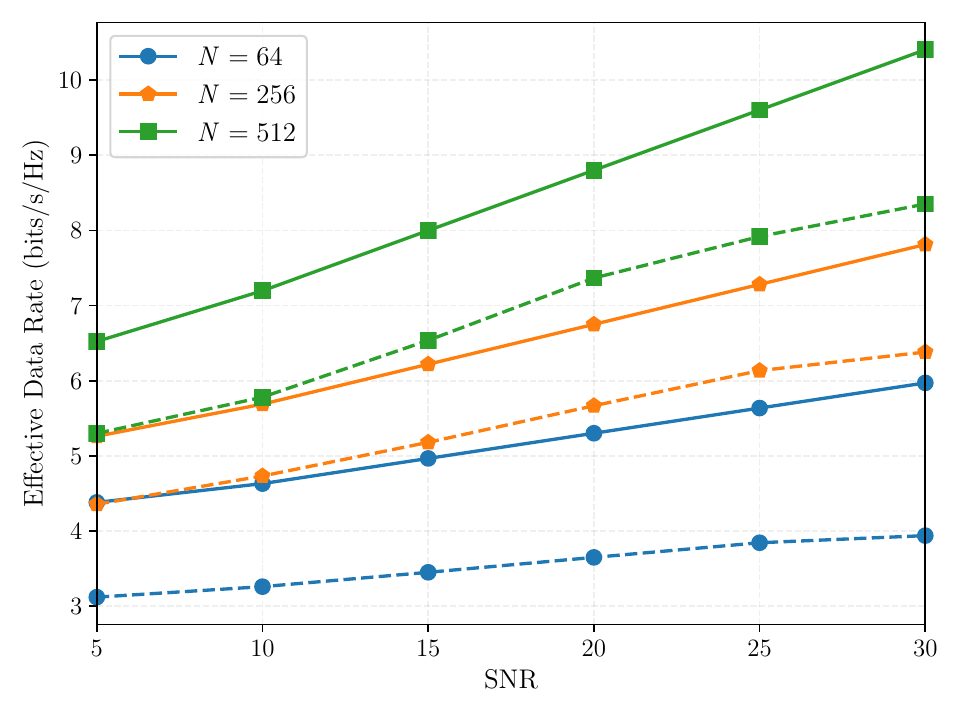}
    \caption{Average data rate versus SNR under different RIS elements. Solid line: DEDT, Dashed line: RCDT}
    \label{reward_via_snr}
\end{figure}

Fig.~\ref{reward_via_snr} compares the effective data rates of the proposed DEDT method and the baseline RCDT method across varying SNR levels and different numbers of RIS elements, with both methods evaluated at their respective optimal mask ratios. As shown in the figure, the effective data rate increases with the SNR, which is expected due to the improvement in signal quality at higher SNR levels. Notably, DEDT consistently outperforms RCDT across all conditions, indicating the superior performance of DEDT. This demonstrates that the proposed DEDT method, by leveraging the DM network to generate channels, is capable of effectively inferring global channel information from a subset of RIS elements. This capability allows DEDT to make more accurate beamforming decisions, resulting in higher effective data rates compared to RCDT.

\begin{figure}[htbp]
    \centering
    \includegraphics[width=0.45\textwidth]{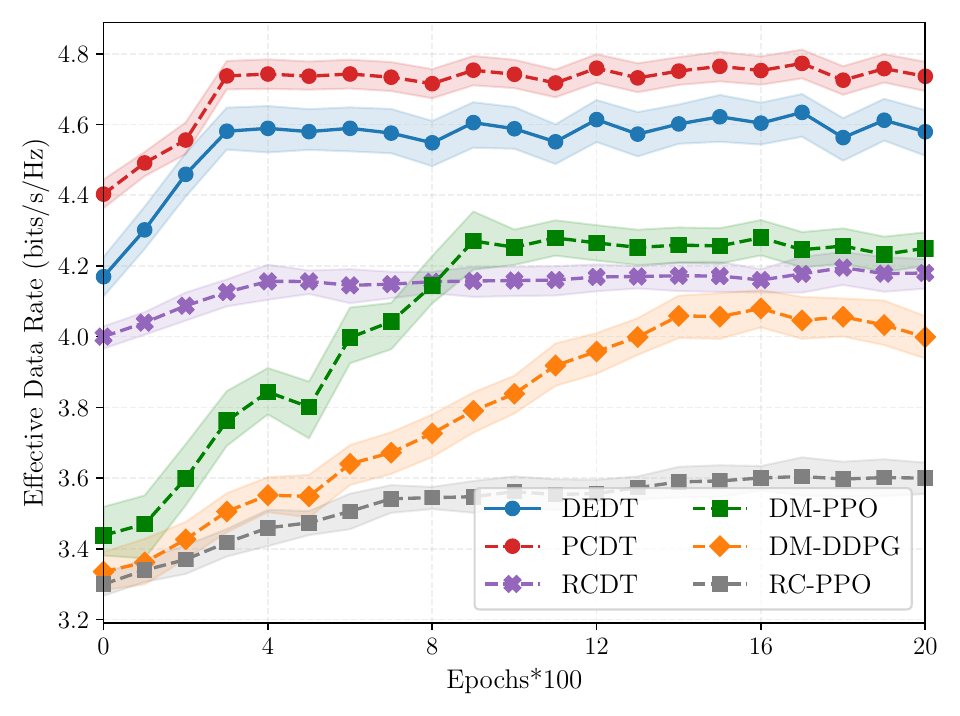}
    \caption{Performance and convergence speed comparison between DEDT and other benchmarks with $64$ elements under $5$ $\mathrm{dB}$ SNR.}
    \label{reward_via_algorithms}
\end{figure}

\begin{figure}[htbp]
    \centering
    \includegraphics[width=0.45\textwidth]{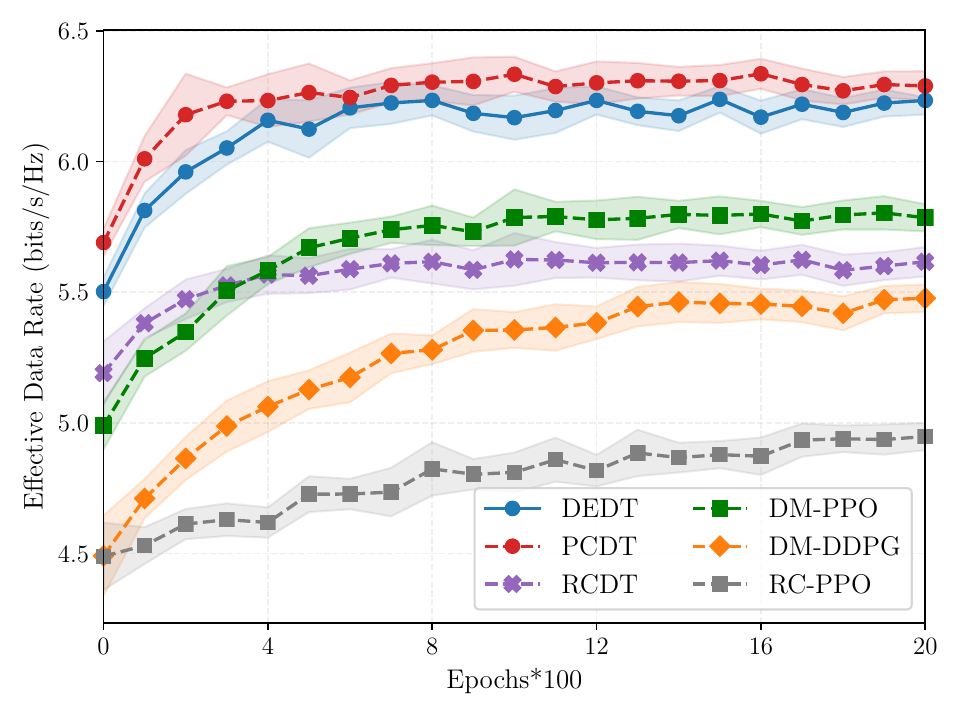}
    \caption{Performance and convergence speed comparison between DEDT and other benchmarks with $256$ elements under $15$ $\mathrm{dB}$ SNR.}
    \label{reward_via_algorithms_256_15dB}
\end{figure}

\begin{figure}[htbp]
    \centering
    \includegraphics[width=0.45\textwidth]{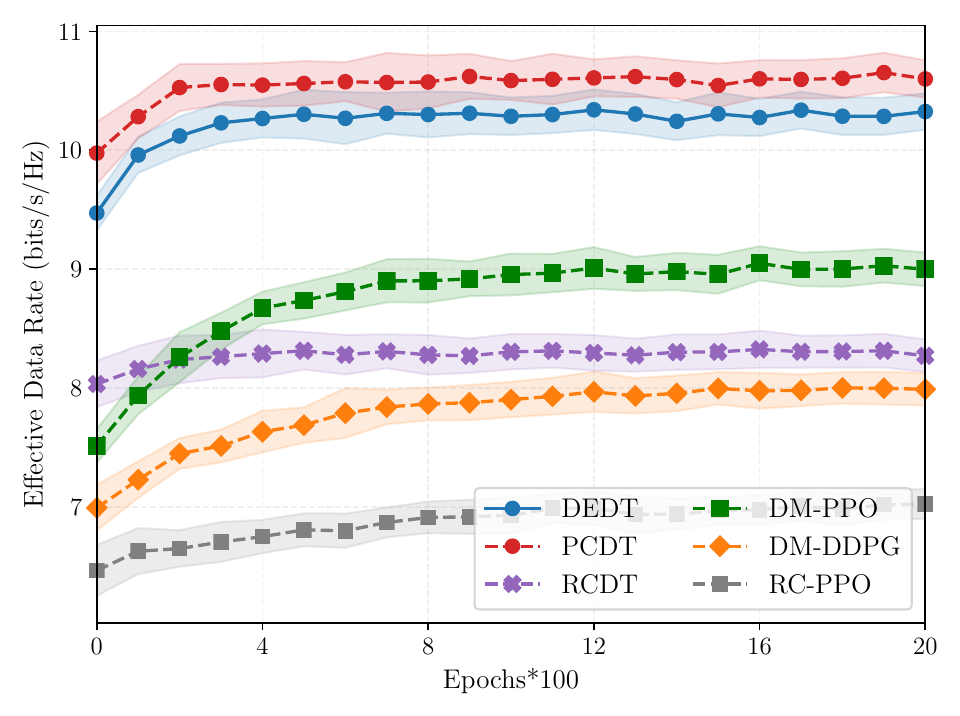}
    \caption{Performance and convergence speed comparison between DEDT and other benchmarks with $512$ elements under $30$ $\mathrm{dB}$ SNR.}
    \label{reward_via_algorithms_512_30dB}
\end{figure}

Figs.~\ref{reward_via_algorithms}-\ref{reward_via_algorithms_512_30dB} present the convergence curves of the effective data rate for various methods under different RIS element configurations and SNR levels. Three figures represent the performance for $64$ elements at $5$ dB, $256$ elements at $15$ dB, and $512$ elements at $30$ dB, respectively. The comparison includes the following methods:
\begin{itemize}
\item [$\bullet$] \textbf{PCDT}: Using a perfect channel as an upper bound for performance.
\item [$\bullet$] \textbf{RCDT}: Random channel generation followed by decision-making using a DT.
\item [$\bullet$] \textbf{DM-PPO}: Channel generation with the DM model, followed by training using Proximal Policy Optimization (PPO) algorithm in RL.
\item [$\bullet$] \textbf{DM-DDPG}: Channel generation with the DM model, followed by training using Deep Deterministic Policy Gradient (DDPG) in RL.
\item [$\bullet$] \textbf{RC-PPO}: Random channel generation followed by training using PPO.
\end{itemize}

These three figures consistently highlights that the proposed DEDT converges much faster than the RL-based methods, reaching convergence in approximately $300$ epochs, while RL-based  approaches take more than $900$ epochs to converge. This represents a speedup of over $3$ times. Furthermore, the performance of DEDT surpasses RL methods by approximately $7.5\%$, which can be attributed to the use of the Transformer-based model in DEDT. The Transformer’s self-attention mechanism does not rely on the approximation of value functions but directly learns the mapping between trajectory sequences and near-optimal actions. Additionally, when comparing DEDT with PCDT, we observe that DEDT closely approaches the upper bound performance achieved with a perfect channel, further demonstrating the high-quality channel generation capability of our method. The three figures together demonstrate that the proposed DEDT method can quickly generalize and converge and achieve near-optimal performance in new environments under diverse conditions with significantly reduced training time.


\section{Conclusions}
In this paper, we have introduced a DEDT framework for optimizing beamforming strategies in RIS-assisted communication systems. By integrating a DM for efficient CSI acquisition with a DT for dynamic beamforming optimization, our method significantly enhances both the accuracy of channel acquisition and the adaptability of beamforming strategies to varying environments. DEDT uniquely leverages the spatial correlations inherent in wireless channels, enabling rapid and accurate derivation of full CSI from partial CSI. This capability reduces the pilot overhead, thereby maximizing the effective use of available spectral resources. Furthermore, the DT component of our framework learns from historical data to make decisions without the need for extensive retraining. This results in adaptable strategies that can swiftly adjust to new environments and maintain near-optimal performance. Simulation results have shown that our method offers improved performance compared to RL algorithms, showcasing enhancements in performance and convergence.

\appendix
\subsection{Proof of Lemma 1}\label{proof of lemma 1}
The element in the $i$-th row and $j$-th column of the spatial correlation matrix is given by
\begin{equation}
    \boldsymbol{\mathrm{R}}^{i,j}=\mathbb{E}\left[e^{j\boldsymbol{\mathrm{q}}(\alpha,\beta)^{T}(\boldsymbol{\mathrm{p}}(n_1,n_2)-\boldsymbol{\mathrm{p}}(m_1,m_2))}\right].
\end{equation}
It can be expanded as 
\begin{equation}\label{element of the spatial correlation matrix}
\begin{aligned}
    \boldsymbol{\mathrm{R}}^{i,j}
    =\mathbb{E}\left[e^{j\frac{2\pi}{\lambda}\left[(n_1-m_1)d_1\sin{(\alpha)}\cos{(\beta)}+(n_2-m_2)d_2\sin{(\beta)}\right]}\right].
\end{aligned}
\end{equation}
Let $C_{1}=\frac{2\pi}{\lambda}(n_1-m_1)d_1$ and $C_{2}=\frac{2\pi}{\lambda}(n_2-m_2)d_2$, then (\ref{element of the spatial correlation matrix}) can be expressed as 
\begin{equation}\label{element of the spatial correlation matrix 2}
\begin{aligned}
    &\boldsymbol{\mathrm{R}}^{i,j}=\int_{-\frac{\pi}{2}}^{\frac{\pi}{2}}\int_{-\frac{\pi}{2}}^{\frac{\pi}{2}}e^{jC_{1}\sin{(\alpha)}\cos{(\beta)}}e^{jC_{2}\sin{(\beta)}}f(\alpha,\beta)\mathrm{d}\alpha\mathrm{d}\beta\\
    &=\int_{-\frac{\pi}{2}}^{\frac{\pi}{2}}e^{jC_{2}\sin{(\beta)}}\frac{\cos{(\beta)}}{2\pi}\mathrm{d}\beta\int_{-\frac{\pi}{2}}^{\frac{\pi}{2}}e^{jC_{1}\cos{(\beta)}\sin{(\alpha)}}\mathrm{d}\alpha.
\end{aligned}
\end{equation}
We first focus on the integration over the azimuth angle $\alpha$, which can be expanded according to Euler's formula as
\begin{equation}\label{integration of the alpha}
\begin{aligned}
    \int_{-\frac{\pi}{2}}^{\frac{\pi}{2}}e^{jC_{1}\cos{(\beta)}\sin{(\alpha)}}\mathrm{d}\alpha=\int_{-\frac{\pi}{2}}^{\frac{\pi}{2}}\cos{(C_{1}\cos{(\beta)}\sin{(\alpha)})}\mathrm{d}\alpha,
\end{aligned}
\end{equation}
where the sine function component is omitted because the sine function is an odd function, and its integral over the symmetric interval from $-\frac{\pi}{2}$ to $\frac{\pi}{2}$ is zero. According to the Jacobi-Anger expansion, the integrand in (\ref{integration of the alpha}) can be derived as 
\begin{equation}
    J_{0}(C_{1}\cos{(\beta)})+2\sum_{n=1}^{\infty}J_{2n}(C_{1}\cos{(\beta)})\cos{(2n\alpha)},
\end{equation}
where $J_{n}(x)=\sum_{m=0}^{\infty}\frac{(-1)^{m}}{m!\Gamma(m+n+1)}\left(\frac{x}{2}\right)^{2m+n}$ is the $n$-th Bessel function and $\Gamma(x)$ is the gamma function. Notice that the integral of $\cos{(2n\alpha)}$ in $-\frac{\pi}{2}$ to $\frac{\pi}{2}$ equals zero. Thus, (\ref{element of the spatial correlation matrix 2}) can be simplified as 
\begin{equation}\label{element of the spatial correlation matrix 3}
\begin{aligned}
    &\boldsymbol{\mathrm{R}}^{i,j}=\frac{1}{2}\int_{-\frac{\pi}{2}}^{\frac{\pi}{2}}e^{jC_{2}\sin{(\beta)}}\cos{(\beta)}J_{0}(C_{1}\cos{(\beta)})\mathrm{d}\beta\\
    &=\frac{1}{2}\int_{-\frac{\pi}{2}}^{\frac{\pi}{2}}e^{jC_{2}\sin{(\beta)}}\cos{(\beta)}\sum_{m=0}^{\infty}\frac{(-1)^{m}(C_1\cos{(\beta)})^{2m}}{4^{m}\Gamma(m+1)m!}\mathrm{d}\beta\\
    &=\frac{1}{2}\sum_{m=0}^{\infty}\frac{(-1)^{m}(C_1)^{2m}}{4^{m}\Gamma(m+1)m!}\int_{-\frac{\pi}{2}}^{\frac{\pi}{2}}\cos{(C_{2}\sin{(\beta)})}\cos^{2m+1}{(\beta)}\mathrm{d}\beta.
\end{aligned}
\end{equation}
When $m\neq 0$ and $C_{2}\neq 0$, the integral part of (\ref{element of the spatial correlation matrix 3}) can be simplified to
\begin{equation}
\begin{aligned}\label{element of the spatial correlation matrix m not equal 0}
    &\int_{-\frac{\pi}{2}}^{\frac{\pi}{2}}\cos{(C_{2}\sin{(\beta)})}\cos^{2m+1}{(\beta)}\mathrm{d}\beta\\
    &=\frac{1}{C_{2}}\cos^{2m}{(\beta)}\sin{(C_2\sin{(\beta)})}\Big|_{-\frac{\pi}{2}}^{\frac{\pi}{2}}\\
    &\quad-\frac{2m}{C_{2}}\int_{-\frac{\pi}{2}}^{\frac{\pi}{2}}\sin{(C_2\sin{(\beta)})}\cos^{2m-1}{(\beta)}\mathrm{d}\beta=0.
\end{aligned}
\end{equation}
Therefore, we only need to consider the part where $m=0$ and $C_{2}\neq 0$. When $m=0$, the gamma function equals $1$, and (\ref{element of the spatial correlation matrix 3}) can be further rewritten as
\begin{equation}
\begin{aligned}
    \boldsymbol{\mathrm{R}}^{i,j}&=\frac{1}{2}\int_{-\frac{\pi}{2}}^{\frac{\pi}{2}}\cos{(C_{2}\sin{(\beta)})}\cos{(\beta)}\mathrm{d}\beta\\
    &=\frac{1}{2C_2}\int_{-\frac{\pi}{2}}^{\frac{\pi}{2}}\mathrm{d}\sin{(C_2\sin{\beta})}\\
    &=\frac{\sin{(C_2)}}{C_2}=\mathrm{sinc}\left(\frac{2\pi d_2}{\lambda}(n_2-m_2)\right).
\end{aligned}
\end{equation}

\subsection{Proof of the Diffusion Loss Function}\label{proof of diffusion loss function}
We begin with the cross entropy in (\ref{cross_entropy})
\begin{equation}\label{cross_entropy_expand_1}
\begin{aligned}
    &L_{\mathrm{CE}}=-\mathbb{E}_{q(\boldsymbol{x}^{(0)}_{t})}\log\left(q_{\theta}(\boldsymbol{x}^{(0)}_{t}|\boldsymbol{x}_{t}^{\mathrm{p}})\right)\\
    &\overset{(a)}\leq-\mathbb{E}_{q(\boldsymbol{x}^{(0)}_{t})}\mathbb{E}_{q(\boldsymbol{x}^{(1:K)}_{t})}\log\left(\frac{q_{\theta}(\boldsymbol{x}_{t}^{(0:K)}|\boldsymbol{x}_{t}^{\mathrm{p}})}{q(\boldsymbol{x}_{t}^{1:K})}\right)\\
    &=\mathbb{E}_{q}\left[-\log q_{\theta}(\boldsymbol{x}_{t}^{(K)}|\boldsymbol{x}_{t}^{\mathrm{p}})+\sum_{k=2}^{K}\log q(\boldsymbol{x}_{t}^{(k)}|\boldsymbol{x}_{t}^{(k-1)})\right.\\
    &\left.\quad -\sum_{k=2}^{K}\log q_{\theta}(\boldsymbol{x}_{t}^{(k-1)}|\boldsymbol{x}_{t}^{(k)},\boldsymbol{x}_{t}^{\mathrm{p}})+\log\frac{q(\boldsymbol{x}_{t}^{(1)}|\boldsymbol{x}_{t}^{(0)})}{q_{\theta}(\boldsymbol{x}_{t}^{(0)}|\boldsymbol{x}_{t}^{(1)},\boldsymbol{x}_{t}^{\mathrm{p}})} \right],
\end{aligned}
\end{equation}
where $\boldsymbol{x}_{t}^{0:K}=(\boldsymbol{x}_{t}^{(0)},\dots,\boldsymbol{x}_{t}^{(K)})$ and the inequality (a) holds due to the convex property of the logarithmic function. Notice that $q(\boldsymbol{x}_{t}^{(k)}|\boldsymbol{x}_{t}^{(k-1)})$ can be rewritten according to the Bayes formula conditioned on the original data $\boldsymbol{x}_{t}^{(0)}$, then (\ref{cross_entropy_expand_1}) can be further express as
\begin{equation}
\begin{aligned}
    &\mathbb{E}_{q}\left[-\log q_{\theta}(\boldsymbol{x}_{t}^{(K)}|\boldsymbol{x}_{t}^{\mathrm{p}})+\sum_{k=2}^{K}\log \frac{q(\boldsymbol{x}_{t}^{(k-1)}|\boldsymbol{x}_{t}^{(k)},\boldsymbol{x}_{t}^{(0)})}{q_{\theta}(\boldsymbol{x}_{t}^{(k-1)}|\boldsymbol{x}_{t}^{(k)},\boldsymbol{x}_{t}^{\mathrm{p}})}\right.\\
    &\quad\left.+\sum_{k=2}^{K}\log \frac{q(\boldsymbol{x}_{t}^{(k)}|\boldsymbol{x}_{t}^{(0)})}{q(\boldsymbol{x}_{t}^{(k-1)}|\boldsymbol{x}_{t}^{(0)})}+\log\frac{q(\boldsymbol{x}_{t}^{(1)}|\boldsymbol{x}_{t}^{(0)})}{q_{\theta}(\boldsymbol{x}_{t}^{(0)}|\boldsymbol{x}_{t}^{(1)},\boldsymbol{x}_{t}^{\mathrm{p}})} \right]\\
    &=\mathbb{E}_{q}\left[D_{\mathrm{KL}}(q(\boldsymbol{x}_{t}^{(K)}|\boldsymbol{x}_{t}^{(0)})||q_{\theta}(\boldsymbol{x}_{t}^{(K)}|\boldsymbol{x}_{t}^{\mathrm{p}}))\right.\\
    &\left.\quad+ \sum_{k=2}^{K}D_{\mathrm{KL}}(q(\boldsymbol{x}_{t}^{(k-1)}|\boldsymbol{x}_{t}^{(k)},\boldsymbol{x}_{t}^{(0)})||q_{\theta}(\boldsymbol{x}_{t}^{(k-1)}|\boldsymbol{x}_{t}^{(k)},\boldsymbol{x}_{t}^{\mathrm{p}}))\right.\\
    &\left.\quad-\log q_{\theta}(\boldsymbol{x}_{t}^{(0)}|\boldsymbol{x}_{t}^{(1)},\boldsymbol{x}_{t}^{\mathrm{p}}) \right],\\
    &=L_{0}+\sum_{k=1}^{K-1}L_{1,k}+L_{2},    
\end{aligned}
\end{equation}
where $D_{\mathrm{KL}}(x||y)=\mathbb{E}_{x}\left[\log(\frac{x}{y})\right]$ is the KL divergence. The KL divergence of two Gaussian distributions can be used to obtain their closed form solutions. Therefore, for the loss function term $L_{1,k}$, we can express it as
\begin{equation}
\begin{aligned}
    L_{1,k}=\mathbb{E}\left[\frac{b_{k}^{2}}{2(1-b_{k})(1-\bar{a}_{k})}||\epsilon_{t}^{(k)}-\tilde{\epsilon}_{\theta,t}^{(k)}(\boldsymbol{x}_{t}^{(k)},\boldsymbol{x}_{t}^{\mathrm{p}},t)||^{2}\right].
\end{aligned}
\end{equation}
The coefficient part of $L_{1,k}$ can be omitted, thus the loss function can be further expressed as
\begin{equation}
    L_{1,k}=\mathbb{E}\left[||\boldsymbol{\epsilon}_{t}^{(k)}-\tilde{\boldsymbol{\epsilon}}_{\theta,t}^{(k)}(\boldsymbol{x}_{t}^{(k)},\boldsymbol{x}_{t}^{\mathrm{p}},k)||^{2}\right].
\end{equation}

\bibliographystyle{IEEEtran}
\bibliography{Ref}

\end{document}